\documentclass[prb,twocolumn,amssymb,floatfix,eqsecnum,showpacs]{revtex4}
\usepackage{graphicx}
\usepackage{bm}
\usepackage{amsmath}
\begin{document}

\title{Theory of the NMR relaxation rates
in cuprate superconductors with field induced antiferromagnetic order}

\author{Yan Chen,$^1$ Jian-Xin Zhu,$^2$ and C. S. Ting$^1$}
\affiliation{$^1$Texas Center for Superconductivity and Department of Physics,
University of Houston, Houston, TX 77204\\ $^2$Theoretical Division, MS B262,
Los Alamos National Laboratory, Los Alamos, NM 87545}

\date{\today}

\begin{abstract}
Based on a model Hamiltonian with a $d$-wave pairing interaction
and a competing antiferromagnetic interaction, we numerically
study the site dependence of the nuclear spin resonance (NMR)
relaxation rate $T_1^{-1}$ as a function of temperature for a
$d$-wave superconductor(DSC) with magnetic field induced spin
density wave (SDW) order. In the presence of the induced SDW, we
find that there exists no simple direct relationship between NMR
signal rate $T_1^{-1}$ and low energy local density of states
while these two quantities are linearly proportional to each other
in a pure DSC. In the vortex core region, $T_1^{-1}$ on $^{17}O$
site may exhibit a double-peak behavior, one sharp and one broad,
as the temperature is increased to the superconductivity
transition temperature $T_c$, in contrast to a single broad peak
for a pure DSC. The existence of the sharp peak corresponds to the
disappearance of the induced SDW above a certain temperature
$T_{AF}$ which is assumed to be considerably lower than $T_c$. We
also show the differences between $T_1^{-1}$ on $^{17}O$ and that
on $^{63}Cu$ as a function of lattice site at different
temperatures and magnetic fields. Our results obtained from the
scenario of the vortex with induced SDW is consistent with recent
NMR and scanning tunneling microscopy experiments.
\end{abstract}


\maketitle

\narrowtext

\section{introduction}

Intensive efforts have been focused recently on the nature of
low-lying excitation spectra around a vortex core in high
temperature superconductors (HTS). The excitations around the core
play a fundamental role in determining physical property of a
superconductor. It has now been established both
experimentally~\cite{Lake01,Miller,NMR,NMR1,NMR2,NMR3} and
theoretically~\cite{Arovas97,Demler,Ogata99,Machida,Zhu01,chen01,Zhu02,Chen02,DChen,Franz}
that a spin density wave (SDW) order could be induced and pinned
at the vortex lattice by a strong magnetic field for both under-
and optimally- doped HTS. In neutron scattering experiments by
Lake {\em et al.}~\cite{Lake01}, a remarkable antiferromagnetic
(AF)-like SDW was observed in optimally- and under-doped
La$_{2-x}$Sr$_x$CuO$_4$ under a strong magnetic field. The muon
spin rotation measurement by Miller {\em et al.}~\cite{Miller}
studied the internal magnetic field distribution in the vortex
core of underdoped YBa$_2$Cu$_3$O$_{6+ x}$, and it was revealed a
feature in the high-field tail which fits well to a model with
static alternating magnetization. Recent nuclear magnetic
resonance (NMR) measurements by Mitrovic {\em et al.}~\cite{NMR}
studied the spatially resolved NMR signal in the mixed state of
optimally doped YBa$_2$Cu$_3$O$_{7- \delta}$, and they found
strong AF fluctuations outside the cores and rather different
electronic states inside the vortex cores. Another NMR experiment
by Kakuyanagi {\em et al.}~\cite{NMR3} investigated the magnetism
in and around the vortex core of nearly optimally-doped
Tl$_2$Ba$_2$CuO$_{6+\delta}$, the NMR signal rate $T_1^{-1}$ at Tl
site provides a direct evidence that the AF spin correlation is
significantly enhanced in the vortex core region. From theoretical
point of view, various
approaches~\cite{Arovas97,Demler,Ogata99,Machida,Zhu01} suggest
the existence of an induced AF order inside the core. Some recent
studies~\cite{chen01,Zhu02,Chen02,DChen,Franz} also reveal that
the AF order could propagate outside the vortex cores and form a
SDW.

As proposed in several articles~\cite{Taki99,wortis,Morr}, that
the spatially imaging NMR experiment can be a powerful tool to
investigate the exotic electronic structure around the vortex
cores. The frequency dependence of NMR signal rate reflects the
internal magnetic field distribution and the electronic and
magnetic excitations in the mixed state can be probed through the
temperature and site -dependence of $T_1^{-1}$. The spatial
variation of the vortex lattice in the NMR experiments can be
resolved by the distribution of internal magnetic field. In the
present paper, we study the NMR theory for HTS with the effect of
the magnetic field induced SDW being taken into account. Our
approach is based upon an effective model Hamiltonian with a
$d$-wave pairing interaction $V_{DSC}$ between nearest neighboring
sites and a competing onsite Coulomb repulsion $U$ which may
generate a competing AF order. The parameters are chosen in such a
way that only $d$-wave superconductivity (DSC) prevails in the
optimally doped HTS samples when the magnetic field B=0. Using the
Bogoliubov-de Gennes (BdG) equations, The dynamic spin-spin
correlation function between site i and site j can be numerically
obtained. From this correlation function and the established
methods~\cite{Shastry}, we are able to derive the NMR relaxation
rates $T_1^{-1}$ on $^{17}O$ and on $^{63}Cu$ nuclei, and from
which the NMR signals as a function of temperature T and site
location~\cite{chenAPS} can be determined. Our calculation is
performed for an optimally doped HTS sample with the chosen $U$ so
that the B induced SDW would appear only below a critical
temperature $T_{AF}$ which is considerably smaller than the
superconductivity transition temperature $T_c$. We find that there
exists no simple direct relationship between the NMR signal rate
$T_1^{-1}$ and the low energy local density of states (LDOS) while
these two quantities are linearly proportional to each other in a
pure $d$-wave superconductor ($U=0$). In the core region, we find
that $T_1^{-1}$ exhibits a double-peak behavior, one sharp and one
broad, as the temperature is increased to $T_c$, in contrast to a
single broad peak for a pure $d$-wave superconductor. It is found
that the low temperature peak becomes sharper and moves to lower T
when $T_{AF}$ becomes lower. This result is consistent with the
NMR experiments~\cite{NMR,NMR1,NMR2,NMR3}. We also show the
differences between $T_1^{-1}$ on $^{17}O$ and that on $^{63}Cu$
as a function of lattice site at different temperatures and
magnetic fields. In general the NMR signal at the $^{63}Cu$ site
is larger than that at $^{17}O$ site, and its magnitude can be
quite enhanced at higher magnetic field. In section II we will
outline our method and the numerical scheme for calculating the
NMR relaxation rate. In section III, we will present our numerical
results and their comparison with the experimental measurements.
In addition we are also going to compare our work with other
similar theoretically studies ~\cite{Taki99,Knapp,Taki02}, there
the articles~\cite{Taki99,Knapp} only consider the case for a pure
$d$-wave superconductor. In all these
works~\cite{Taki99,Knapp,Taki02}, the NMR signals are calculated
for the square lattice sites which may not rigorously related to
the $^{17}O$ and $^{63}Cu$ sites as discussed in the present
paper. In the last section, we will give a summary and discussion
of our results.

\section{Formalism}

Let us begin with a phenomenological model in which interactions
describing both DSC and SDW order parameters in a two-dimensional
lattice are considered. The effective Hamiltonian can be written as:
\begin{eqnarray}
H&=&-\sum_{i,j,\sigma} t_{ij} c_{i\sigma}^{\dagger}c_{j\sigma}
+\sum_{i,\sigma}( U n_{i {\bar {\sigma}}} -\mu)
c_{i\sigma}^{\dagger} c_{i\sigma}
\nonumber \\
&&+\sum_{i,j} ( {\Delta_{ij}} c_{i\uparrow}^{\dagger}
c_{j\downarrow}^{\dagger} + h.c.)\;,
\end{eqnarray}
where $c_{i\sigma}^{\dagger}$ is the electron creation operator,
$\mu$ is the chemical potential, and the summation is over the
nearest neighboring sites. In the presence of magnetic field $B$
normal to the two dimensional plane, the hopping integral can be
expressed as $ t_{ij}= t_0 \exp[{i \frac{ \pi}{\Phi_{0}}
\int_{{\bf r}_{j}}^{{\bf r}_{i}} {\bf A}({\bf r})\cdot d{\bf r}}]$
for the nearest neighboring sites $(i,j)$, with $\Phi_0=h/2e$ as
the superconducting flux quantum. We assume that the strength of B
is large enough such that it can be regarded as uniform. Here we
choose the Landau gauge for the vector potential ${\bf
A}=(-By,0,0)$. Since the internal magnetic field induced by the
supercurrent around the vortex core, as it will be numerically
shown later, is so small as compared with B, the above assumption
should be justified. The induced SDW and the DSC orders in our
system are respectively defined as $\Delta^{SDW}_{i} = U \langle
c_{i \uparrow}^{\dagger} c_{i \uparrow} -c_{i
\downarrow}^{\dagger}c_{i \downarrow} \rangle$ and
$\Delta_{ij}=V_{DSC} \langle c_{i\uparrow}c_{j\downarrow}-c_{I
\downarrow} c_{j\uparrow} \rangle /2$, where $U$ and $V_{DSC}$
represent the interaction strengths for SDW and DSC. The
mean-field Hamiltonian (1) can be diagonalized by solving the
resulting BdG equations self-consistently
\begin{equation}
\sum_{j} \left(\begin{array}{cc}
{\cal H}_{ij,\sigma} & \Delta_{ij} \\
\Delta_{ij}^{*} & -{\cal H}_{ij,\bar{\sigma}}^{*}
\end{array}
\right)
\left(\begin{array}{c} u_{j,\sigma}^{n} \\
v_{j,\bar{\sigma}}^{n}
\end{array}
\right) =E_{n} \left(
\begin{array}{c}
u_{i,\sigma}^{n} \\
v_{i,\bar{\sigma}}^{n}
\end{array}
\right)\;,
\end{equation}
where the single particle Hamiltonian ${\cal H}_{ij,\sigma}= -t_{ij}
+(U n_{i \bar{\sigma}} -\mu)\delta_{ij}$, and $n_{i \uparrow} =
\sum_{n} |u_{i\uparrow}^{n}|^2 f(E_{n})$, $ n_{i \downarrow} =
\sum_{n} |v_{i\downarrow}^{n}|^2 ( 1- f(E_{n}))$, $
\Delta_{ij} = \frac{V_{DSC}} {4} \sum_{n} (u_{i\uparrow}^{n}
v_{j\downarrow}^{n*} +v_{i\downarrow}^{*} u_{j\uparrow}^{n})
\tanh \left( \frac{E_{n}} {2k_{B}T} \right)$, with $f(E)$ as
the Fermi distribution function and the electron density $n_{i}=
n_{i \uparrow} + n_{i \downarrow}$. The DSC order parameter at the
$i$th site is $\Delta^{D}_{i}= (\Delta^{D}_{i+e_x,i} + \Delta^{D}_{i-e_x,i}
- \Delta^{D}_{i,i+e_y} -\Delta^{D}_{ i,i-e_y})/4$ where $ \Delta^{D}_{ij} =
\Delta_{ij} \exp[ i { \frac{\pi}{\Phi_{0}} \int_{{\bf r}_{i}}^{({\bf r}_{i}+{\bf r}_{j})/2 }
{\bf A}({\bf r}) \cdot d{\bf r}}]$ and ${\bf e}_{x,y}$
denotes the unit vector along $(x,y)$ direction. The main
procedure of our self-consistent calculation is summarized as follows. For a random
set of initial parameters $n_{i \sigma}$ and $\Delta_{ij}$, the Hamiltonian is
numerically diagonalized and the obtained quasiparticle wave functions $u_{i,\sigma}^{n}$
and $v_{i,\sigma}^{n}$ are used to calculate the new parameters for the next iteration step.
The iteration continues until the relative difference of order parameter between two consecutive
iteration steps is less than $10^{-4}$.

In the following section, we need to calculate the local density of states (LDOS)
according to the formula:
\begin{equation}
\rho_{\bf i}(E) = -\sum_{n}[\vert u_{{\bf
i},\uparrow}^{n}\vert^{2} f^{\prime}(E_{n}-E) +\vert v_{{\bf
i},\downarrow}^{n}\vert^{2}f^{\prime}(E_{n}+E)]\;,
\end{equation}
where $f^{\prime}(E) \equiv df(E)/dE$. The LDOS is proportional to
the local differential tunneling conductance at low temperature
which can be measured by STM experiments. In addition, we also
need to calculate the internal magnetic field distribution of the
vortex state. The internal magnetic field is computed through the
Maxwell equation, $\nabla\times {\bf H}_{int}({\bf
r})={\frac{4\pi}{c}}{\bf j}({\bf r})$, where the current ${\bf
j}({\bf r})$ is calculated as:
\begin{eqnarray}
j_{\bf {e_{x,y}}} ({\bf r}_i) &=& 2 |e| {\rm Im}\{
t_{\bf{i+e_{x,y}}, \bf{i}} \sum_n [ u^{n \ast}_{\bf {i+e_{x,y}}}
u^{n}_{\bf i} f(E_n) \nonumber \\ && + v^{n}_{\bf {i+e_{x,y}}}
v^{n \ast}_{\bf i} (1-f(E_n)) ] \}.
\end{eqnarray}
The nuclear spin relaxation rate $T_1^{-1}$ can be obtained from
the spin-spin correlation function ${\chi _{+,-}(i, j,\Omega)}$ at
zero energy through the following site dependent
function~\cite{Leadon,Taki99},
\begin{eqnarray}
R(i, j) &=&{\rm Im}\chi _{+,-}(i, j,i\Omega _{n}\rightarrow \Omega
+{\rm i}\eta )/(\Omega /T)|_{\Omega \rightarrow 0} \nonumber \\
&=&-\sum_{ n , n^{\prime }} u^{n}_{i} u^{n^{\prime} \ast}_{i} [
u^{n }_{j} u^{n^{\prime} \ast}_{j} + v^{n }_{j} v^{n^{\prime}
\ast}_{j} ] \nonumber \\ &&\times \pi Tf^{\prime }(E_{n})\delta
(E_{n}-E_{n^{\prime}}).
\end{eqnarray}
Since all of the experiments ~\cite{NMR,NMR1,NMR2,NMR3} are
performed at the oxygen, copper, or other nuclei in HTS, we need
the expressions of the NMR relaxation rate at these sites. In this
paper we shall mainly consider the contribution from the planar
oxygen ($^{17}O$) and copper ($^{63}Cu$) nuclei. The oxygen
nucleus is located between two $Cu$ nuclei along the $x$ or $y$
axis. The internal magnetic field $H_{int}$ on $^{17}O$ site is
averaged over its two nearest neighboring $^{63}Cu$ sites. The NMR
relaxation rate form factor on nuclei $^{17}O$ and $^{63}Cu$ in
momentum space is: $F_{O,x} \sim 1+cos(q_x/2)$ and $F_{Cu} \sim
(cos(q_x)+cos(q_y))^{2}$.  It can be expressed in terms of the
real-space spin-spin correlations as follows~\cite{Shastry}:
\begin{eqnarray}
1/^{17}T_1T &=& C [R(i,i)+1/2 \sum_{j,x}^{'} R(i,j)] ,
\\ 1/^{63}T_1T &=& B [R(i,i)+ 1/2 \sum_{j}^{''} R(i,j) \nonumber \\
               &&+1/4\sum_{j}^{'''} R(i,j)] ,
\end{eqnarray}
here the notation of $\sum_{x}^{'}$, $\sum^{''}$ and $\sum^{'''}$
means summation over two nearest-neighboring sites along the $x$
axis, four next-neighboring-sites, and four next-next-neighboring
sites, respectively. Here we calculate $T_1^{-1}$ in terms of
arbitrary unit and choose the ratio of the
coefficients~\cite{Bulut} $B/C=1.583$.

Throughout this paper, we use the following parameters: $U=2.4$,
$V_{DSC}=1.2$, the linear dimension of the unit cell of the vortex
lattice is chosen as $N_{x}\times N_{y}=40\times 20$ sites and the
number of the unit cells $M_{x}\times M_{y}=20\times 40$, and the
hole doping level $x=0.15$. This choice corresponds to a uniform
magnetic field $B \simeq 37 T$. We set $t_0 = a =1$. The standard
procedures~\cite{Wang95} are followed to introduce magnetic unit
cells, where each unit cell contains two superconducting flux
quanta. By introducing the quasi-momentum of the magnetic Bloch
state, we obtain the wave function under the periodic boundary
condition whose region covers many unit cells.

\section{Numerical Results and Comparing with Experiments}

\begin{figure}[b]
\vspace{-0.4in}
\includegraphics[width=15cm]{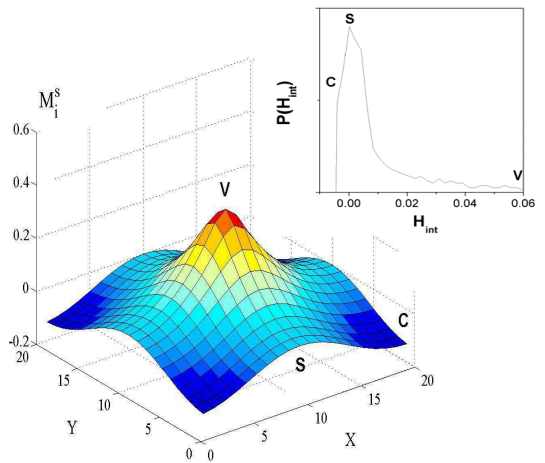}
\vspace{-1.0in} \caption{\label{Fig1} The amplitude distribution
of the SDW order parameter $M_{\bf i}^{s}$ (a) in one magnetic
unit cell. V, S, C points represents the vortex core center, the
saddle point midway between two nearest neighboring vortices, and
the center of the squared vortex lattice unit cell, respectively.
The inset shows the histogram $P(H_{int})$ of the induced magnetic
field distribution where $H_{int}$ is in a unit of $10^{3}$ Gauss.
The calculation is performed with $U=2.4$ and $V_{DSC}=1.2$.}
\end{figure}
Our previous numerical calculations~\cite{chen01} show clearly that the
induced AF order exists both inside and outside the vortex cores, and
 behaves like an inhomogeneous SDW with the same wave length in the $x$ and
 $y$ directions. Fig. 1 depicts the SDW amplitude or the staggered
  magnetization $M_{\bf i}^{s}$=$\Delta^{SDW}_{i}$/$U$
distribution in a square lattice. It is seen that the induced SDW order
reaches its maximum value at the vortex core center (V), zero value at
 the saddle-point midway between two nearest neighboring vortices (S),
 and in between at the center of the squared vortex lattice unit cell (C).
Using Eq.(2.4), the internal magnetic field $H_{int}$ at different
site can be numerically determined. In the upper right inset of
Fig. 1 we show the histogram of $H_{int}$, where $P(H_{int})$
measures the number of lattice sites (or plaquettes) with fixed
$H_{int}$. We can approximately identify each site (V, S, C)
including the sites around it in the vortex lattice to the
distribution. The correspondence between the site position and
$H_{int}$ can roughly be established. It is easy to see that the
magnitude of $H_{int}$ is decreasing along the path
(V$\rightarrow$S$\rightarrow$C) and its direction is always
perpendicular to the lattice plane. In a narrow region, $H_{int}$
may even become negative. This is very different from the case for
a pure DSC where $H_{int}$ is always positive ~\cite{Taki99}. Our
histogram of $H_{int}$ distribution is consistent with the
experimental data~\cite{NMR} except there the region of negative
$H_{int}$ is somewhat wider than ours. The experimentally observed
negative $H_{int}$ may indicate the existence of magnetic field
induced SDW around the vortex cores.

The NMR signal at the maximum cutoff of the histogram as a
function of internal magnetic field comes from the V-site. The
minimum cutoff is from the C-site. The peak value of the internal
magnetic field comes from the S-site. Contrary to the case for a
pure $d$-wave superconductor, the SDW order outside the vortex
core may have a remarkable contribution to both the internal
magnetic distribution and the NMR signal. It is supposed that
$^{17}O$ NMR experiment can provide direct information on the LDOS
as the antiferromagnetism spin fluctuation is
cancelled~\cite{Taki91,Bulut92}. From our study, we find that in
the presence of SDW vortex cores, the internal magnetic field on
$^{17}O$ site can not be cancelled exactly and appreciable residue
exists.

Next, we study the relationship between low energy LDOS and NMR
signal rate $1/T_1$. In a pure $d$-wave superconductor(DSC), a
linear relationship is found between $1/T_1(i) \sim R(i,i)$ and
the LDOS $\rho_{i}(E=0)$~\cite{Taki99} at zero temperature. We
have checked our program and reproduced similar results. Figure 2
depicts the spatial distribution images of LDOS $\rho_{i}(E=0)$
(a), and $1/{T_1(i)T}$ (b) for a pure DSC with $U$=0 at low
temperature. In this case, the core state shows zero-energy
peak~\cite{Wang95} and nodal directions are along $x=y$ and
$x=-y$. The images in Fig. 2(a) and 2(b) are similar to each other
and the results can be regarded as linearly proportional to each
other. Thus the NMR signal for a pure DSC can be estimated
quantitatively from its corresponding zero energy LDOS according
to the linear relationship.

\begin{figure}[t]
\includegraphics[width=8.6cm]{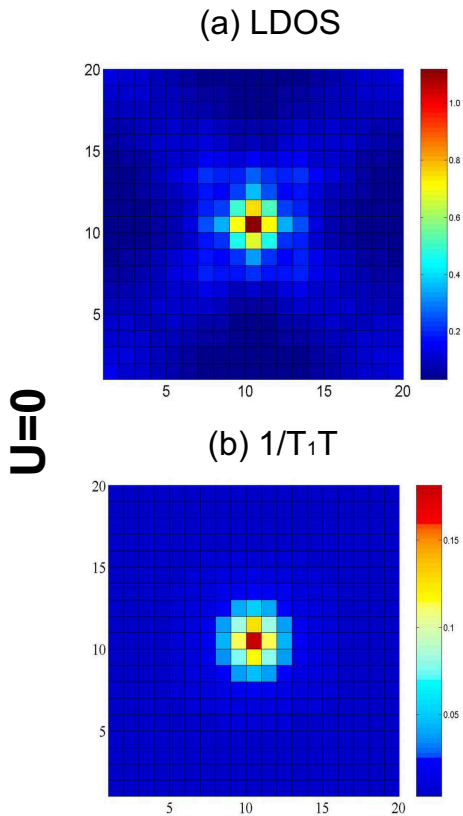}
\caption{\label{Fig2} The spatial distribution images of the LDOS
at the energy $E=0$ (a) and NMR relaxation rate $1/T_1(i)T$ (b) in
a vortex unit cell for a pure DSC with $U$=0. Other parameter
values are the same as those in Fig. 1.}
\end{figure}
From Fig. 2(a), the LDOS of quasiparticles decreases along the
path (V$\rightarrow$C$\rightarrow$S). In the nodal directions,
appreciable weight of the low-energy states extending from the
vortex core still remains there. Our numerical result indicates
that the LDOS of the low-energy states at the C-site is slightly
larger than that of the S-site, although C-site is farther away
from the vortex center. This is different from the work by
Takigawa {\em et al}.~\cite{Taki99} where the LDOS at the S site
is found to be larger than that at the C site. The
authors~\cite{Taki99} there employed the symmetrical gauge and
they found that their unit vector of vortex lattice is oriented
along $45^{o}$ from the Cu-O bond. But our result is consistent
with the study of Knapp {\em et al.} ~\cite{Knapp}who used a
different approach and calculated the NMR signal at various sites
for a pure DSC. Furthermore, a recent small-angle neutron
scattering measurement~\cite{smallangle} for
La$_{1.83}$Sr$_{0.17}$CuO$_{4+\delta}$ indicates that the unit
vector of vortex lattice is clearly oriented along the Cu-O bonds,
consistent with our result in Fig. 1. The NMR
experiments~\cite{NMR,NMR1,NMR2,NMR3} also show that the NMR
signal first decreases as one moves away from the V-site to S-site
and then increases at the C-site, consistent with the results in
Fig. 2. Here we notice that the NMR signal at C site is slightly
larger than that at S-site, their difference could be seen in Fig.
6 and Fig. 7. It appears that the above analysis could provide a
basic understanding of the spatial distribution of the
experimentally observed NMR relaxation rates in HTS in the
framework of pure DSC. However, it is well known that the
theoretical predicted LDOS~\cite{Wang95}, which shows a strong
zero energy peak for a pure DSC at the vortex core is in
contradictory to the STM measurements~\cite{Renner,Pan} on HTS. In
order to overcome this difficulty, the magnetic induced AF order
has been introduced~\cite{Zhu01} to account for the experiments.
In the following we shall investigate the effect of induced AF or
SDW order on the NMR theory in HTS.

The presence of the induced SDW order in the mixed state may lead
to violation of the linear relationship between the LDOS at $E$=0
and the NMR relaxation rate $1/T_1(i)$ . This can be seen clearly
in Fig. 3, where we show the spatial images of the LDOS (a) and
the $1/{T_1(i)T}$ (b) according to our numerical calculations for
$U$=2.4. There is little resemblance between Figs. 3(a) and 3(b),
and the linear relationship does not exists here. Fig. 3(a)
indicates that the LDOS is smallest at V-site, and largest at
S-site, it clearly can not be applied to explain the NMR
experiments~\cite{NMR,NMR1,NMR2,NMR3} as in the case for a pure
DSC .
\begin{figure}[t]
\includegraphics[width=8.6cm]{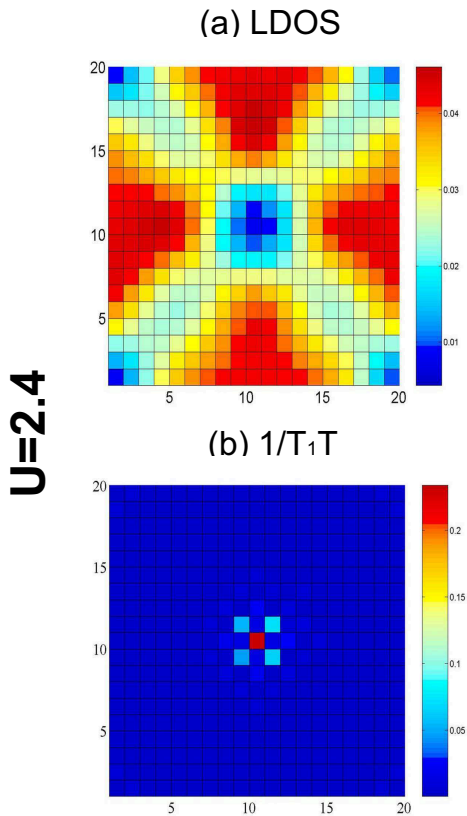}
\caption{\label{Fig3} The spatial distribution images of the LDOS
at the energy $E=0$ (a) and NMR relaxation rate $1/{T_1(i)T}$ (b)
in a vortex unit cell for a DSC with induced SDW.  All parameter
values are the same as those in Fig. 1.}
\end{figure}
But the LDOS obtained here exhibits no zero energy peak at the
vortex core which is consistent with the STM
experiments~\cite{Renner,Pan}. The averaged LDOS of quasiparticles
in Fig. 3(a)($U$=2.4) is greatly suppressed than that in Fig.
2(a)($U$=0). This is because the presence of the induced SDW
enhances the insulating nature of the system and thus reduces the
number of the quasiparticles. When $U$=2.4, there are two
contributions to the NMR signal $1/{T_1(i)T}$, at the low
temperature, the dominant one is from the induced SDW order and
the minor one is from the low energy quasiparticle states. The
LDOS is the smallest at the vortex core center but the SDW order
is at its maximum as shown in Fig. 1. From Fig. 3(b), it is easy
to see that the spatially distributed NMR signal $1/T_1(i)T$ which
has the largest value at the vortex center because of the induced
AF order there and decreases along the path
(V$\rightarrow$C$\rightarrow$S). This feature is consistent with
what has been observed by the
experiments~\cite{NMR,NMR1,NMR2,NMR3} and in sharp contrast with
Takigawa {\em et al.}'s results~\cite{Taki02}. They claimed that
$T_1^{-1}$ at V-site is smaller than the neighboring sites and the
NMR signal rate at S-site is higher than that of C-site. The
choice of lattice size in their calculations corresponds to a very
strong magnetic field at about 51 Tesla. Their results seem to be
inconsistent with the measurement of $1/^{17}T_1$ on
YBa$_2$Cu$_3$O$_{7- \delta}$~\cite{NMR}.

The NMR signals also exhibit the temperature T and magnetic field
B dependences. Our approach described in Section II is easily
extended to study the finite T case, but not the B dependence
because weaker B implies larger size calculation which we are not
able to do at the present moment. Since the observed NMR signals
are coming from the oxygen, copper, or other atoms in HTS, here we
shall mainly consider the temperature dependence of the NMR signal
rate $1/^{17}T_1$ from the $^{17}O$ sites using Eq. (2.6). It
needs to be point out that the value of $R(i,j \neq i)$ in Eq.
(2.6) is of the same order with $R(i,i)$ and its contribution
should not be neglected. In Fig. 4, the NMR relaxation rates
$1/T_1$=$1/^{17}T_1$ on $^{17}O$ site normalized by its value
$1/T_1^{c}$ at $T_c$ as a function of T are plotted for sites V, C
and S. The calculations are performed using $U$=2.4 except the
dashed curve associated with the V site where $U$=0 and is
obtained for a pure DSC. The inset plots the temperature
dependence of the staggered magnetization at V-site and DSC order
at S-site.
\begin{figure}[t]
\includegraphics[width=9.0cm]{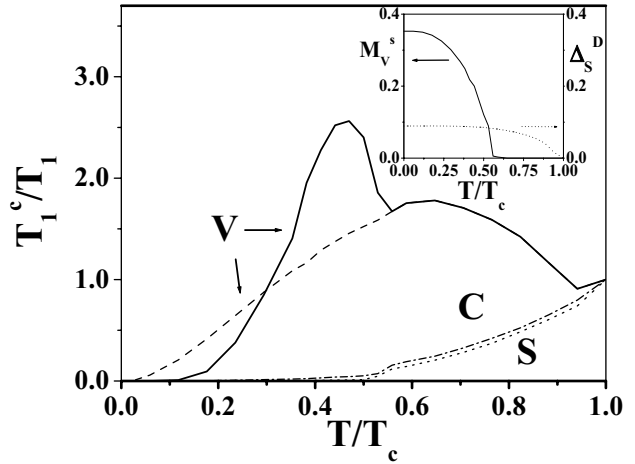}
\caption{\label{Fig4} Temperature dependences of $^{17}O$ NMR
signals $1/T_1$ normalized by its value $1/T_1^c$ at $T_c$ are
plotted as a function of $T/T_c$ at the sites V, S, C.  The dashed
line associated with the site V is obtained for a pure DSC by
using $U$=0. The inset plots the temperature dependence of SDW
order parameters at V-site and DSC order at S-site.}
\end{figure}
Our calculations show that the vast majority sites outside the
vortex core exhibit long relaxation time, but few sites around the
vortex core exhibit short relaxation time, which contains useful
information on the site-dependent low energy excitations around
the vortex. The maximum magnitude of $^{17}T_1^{-1}$ at V-site is
about two orders of magnitude larger than that far from the vortex
core, and the NMR signal rate at C-site is also larger than that
at S-site. Because both NMR signals at C and S sites are very weak
as compared to that at V site, their distinction is not reflected
in Fig. 4 below $T < 0.4T_c$. The minimal value of the staggered
magnetization occurs near the saddle point S. It reflects the fact
that the SDW order at C site is larger than that at S site as
shown in Fig. 1. This result clearly indicates that the presence
of the induced SDW order makes remarkable contribution to NMR
signals. At the vortex core center V, the evolution of $T_1^{-1}$
with temperature exhibits a double peak structure below $T_c$; the
sharp peak occurs at lower T and the broad peak is at higher T.
This is in contrast to a single peak obtained with $U$=0 for a
pure DSC~\cite{Taki99}, and it can be clearly seen from the dashed
curve below a critical temperature $T_{AF} \sim 0.54 T_c$ and the
solid curve above $T_{AF}$ associated with the V site in Fig. 4.
The inset shows that above the temperature $T_{AF}$, the induced
staggered magnetization vanishes and only DSC order prevails up to
$T_c$. The sharp peak at ~0.45$T_c$ is below $T_{AF}$ and  the
broad peak at ~0.64$T_c$ is originated in the pure DSC order. It
is obvious that the value of the critical temperature $T_{AF}$ is
$U$ or sample dependent, and the position of the sharp peak can
made to a lower T if a smaller $U$ or $T_{AF}$ is chosen for the
calculation. In case the value of $T_{AF}$ is slightly higher than
the position of the broad peak, only a single peak for the NMR
signal is predicted. When $T_{AF}$ is close or larger than $T_c$,
the NMR signal would be a increasing function of T and may not
show any peak structure. We also have done the same calculation
for the NMR relaxation rates on $^{63}Cu$ nuclei on our model
lattice sites $(1/T_1(i))$,  and found similar T and site
dependence as shown in Fig. 4. In the experimental aspect, a sharp
peak has been observed at $T \sim 0.235 T_c$ ~\cite{NMR3} in
Tl$_2$Ba$_2$CuO$_6$ while there is no observed peak in
YBa$_2$Cu$_3$O$_7$ up to $T \sim 0.272 T_c$ ~\cite{NMR}. Therefore
the value $U$ in YBa$_2$Cu$_3$O$_7$ should be larger than that in
Tl$_2$Ba$_2$CuO$_6$. This conclusion is opposite to the
theoretical prediction made by Takigawa {\em et al.}~\cite{Taki02}

\begin{figure}[t]
\includegraphics[width=10.2cm]{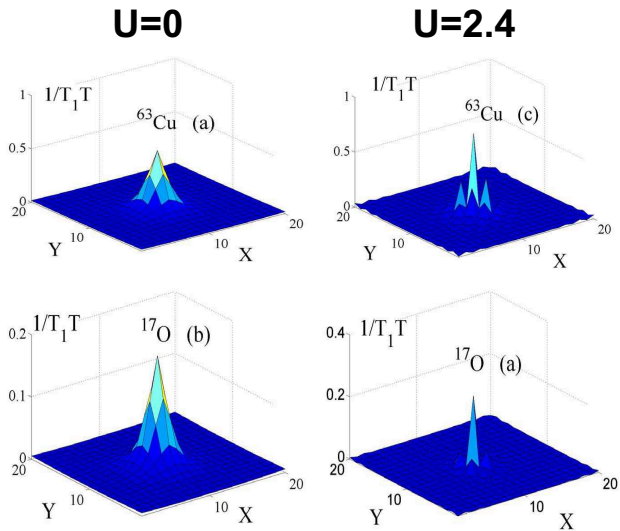}
\caption{\label{Fig5} Site-dependent relaxation rate $T_1^{-1}$
for $^{63}Cu$ (a), $^{17}O$ nuclei of pure $d$-wave case and
$^{63}Cu$ (c), $^{17}O$ (d) site of SDW vortex core case.}
\end{figure}
We also would like to examine the difference in the spatial
distributions of the NMR signals from $^{63}Cu$ and $^{17}O$
nuclei. In Fig. 5, The site distribution maps of $T_1^{-1}$ for
$^{63}Cu$ and for $^{17}O$ in a pure DSC ($U$=0) are respectively
illustrated in (a) and (b). Both of these maps reach a similar
peak value at the vortex core and then fall away smoothly from it.
In the vortex core, the magnitude of $T_1^{-1}$ at the $^{63}Cu$
site is about three times larger than that at the $^{17}O$ site.
In Figs. 5(c) and 5(d), we present the results for $^{63}Cu$ and
$^{17}O$ nuclei in a DSC with induced SDW ($U$=2.4). The
site-dependent spectra for both nuclei show staggered oscillations
in the vortex core, there the NMR signal rate from the $^{17}O$
site is about one fourth of the value from that of $^{63}Cu$ site.
This is because the AF order at the $^{17}O$ sites could be
dramatically weakened by the partial cancellation of the staggered
magnetization between two nearest neighboring sites on the
original lattice. It is useful to point out that the NMR signal
$^{17}T_1^{-1}$ outside the vortex core region is one order of
magnitude smaller than that of $^{63}T_1^{-1}$. In the presence of
the induced SDW, the NMR signal rate is enhanced at the core
center and suppressed away from the vortex core. For both $U$=0
and $U$=2.4 cases, $T_1^{-1}$ reaches a maximum at the core center
while its value at the C-site is always larger than that at the
S-site.

\begin{figure}[t]
\includegraphics[width=10.0cm]{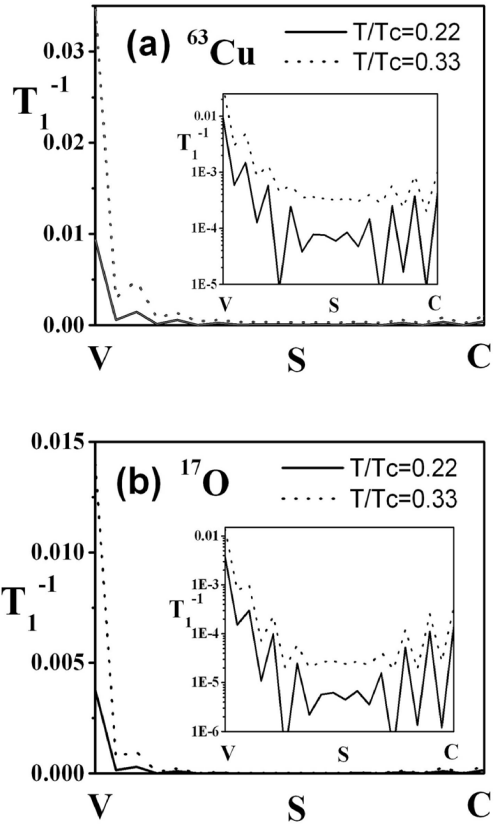}
\vspace{-0.3in} \caption{\label{Fig6} Site-dependent relaxation
rate $T_1^{-1}$ for two different temperatures with $U$=2.4 and
the magnetic field B=37T at $^{63}Cu$ (a) and at $^{17}O$ (b). The
log plot is shown as inset. }
\end{figure}
Next, the temperature dependence of NMR signal rates for the case
of $U$=2.4 is examined. Here, we plot the variation of $T_1^{-1}$
along the path $V \rightarrow S \rightarrow C$ for $^{63}Cu$
nuclei in Fig. 6(a) and for $^{17}O$ nuclei Fig. 6(b) at two
different temperatures. The values of $T_1^{-1}$ in Figs. 6(a) and
6(b) exhibit similar behavior and oscillate from site to site. The
amplitude of the oscillation reflects the underlying staggered
magnetization and is apparently weakened at higher temperature. In
fact, the T dependence of NMR signal rate at three typical points
has been seen from Fig. 3. Figure 6 shows more clearly that NMR
signal rate at each site increases with temperature and its value
at site C is always larger than that at site S. We have reproduced
the power law relation $T_1^{-1} \sim T^3$ for a pure
DSC~\cite{Taki99} at zero magnetic field and at low temperature.
In the presence of vortices with induced SDW order, it should not
be surprised that $T_1^{-1}$ does not follow this $T^3$ law.

\begin{figure}[t]
\includegraphics[width=10.0cm]{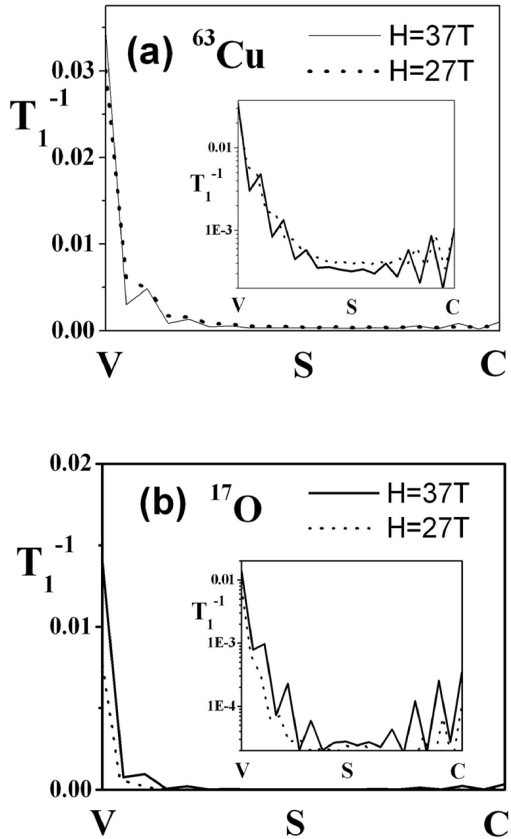}
\vspace{-0.3in} \caption{\label{Fig7} Site-dependent relaxation
time $T_1^{-1}$ for two different magnetic fields with $U$=2.4 and
the temperature $T/T_c$=0.35 at $^{63}Cu$ (a) and at $^{17}O$ (b).
The log plot is shown as inset. }
\end{figure}

 Finally, we show the site dependence of NMR signal
rates on $^{63}Cu$ and on $^{17}O$ nuclei at two different
magnetic field B respectively in Figs. 7(a) and 7(b). The NMR
signals $T_1^{-1}$ in Figs. 7(a) and 7(b) are similar to each
other. Since the magnitude of the induced SDW order is strongly
dependent on the external magnetic field B --- higher B leads to
more pronounced SDW order which may in turn yield a larger NMR
signal. This conclusion is consistent with the NMR
experiments~\cite{NMR3}. As a result, the amplitude of oscillation
at different sites, which measures the strength of the induced
staggered magnetization, is dramatically reduced when the magnetic
field B is decreased from 37T to 27T as shown in Fig. 7. Here
$B=27T$ corresponds to a vortex lattice unit cell of $48 \times
24$ sites. On the other hand, the NMR signal rate at V-site is
suppressed with increasing B for a pure DSC ~\cite{Taki99} which
reflects the density of quasiparticle around the vortex core is
reduced due to the enhancement of quasiparticle leakage along the
nodal directions. It was argued by Takigawa et al.~\cite{Taki02}
that when AF order inside the vortex core is large enough,
$T_1^{-1}$ becomes the absolute minimum at V-site. This
conjecture~\cite{Taki02} is based on the assumption that the
linear relation between $T_1^{-1}$ and LDOS still exists even in
the presence of the SDW order. On the other hand, our results
indicate strongly that with the increasing B, the enhancement of
AF order inside the vortex core may lead to more pronounced
$T_1^{-1}$ value at the V-site. This is very different from what
has been predicted in Ref. 31. Finally we point out that existing
experimental data~\cite{NMR} show the presence of appreciable NMR
signal rate at S-site and its value is insensitive to T. But in
our calculations there is a very weak NMR signal at S-site. The
discrepancy between experimental data and our results may be due
to the reason unclear to us at this moment.

\section{Summary}

Based on a model Hamiltonian as described in Sections I and II, we
have investigated the site- and temperature- dependence of NMR
signal rate $T_1^{-1}$ near the vortex core in an optimally doped
HTS. The calculations are carried out with ($U=2.4$) and without
($U=0$) the magnetic field induced SDW order. Although the
relative values of $T_1^{-1}$ at V, S, and C sites near the vortex
core obtained theoretically for a pure DSC are  consistent with
recent experiments~\cite{NMR,NMR1,NMR2,NMR3}, the temperature
dependent of $T_1^{-1}$ thus derived fails to explain the sharp
peak observed in NMR experiments~\cite{NMR3} for
Tl$_2$Ba$_2$CuO$_{6+\delta}$ at low temperature. Moreover, the
LDOS obtained for a pure DSC ~\cite{Wang95} has a zero energy peak
at the core center and this is in contradictory to the STM
observations~\cite{Renner,Pan}. All these difficulties could
somewhat be removed by introducing the field induced SDW pinned at
the vortex cores as it has been studied in the present paper. With
the induced SDW order, the LDOS exhibits no zero energy peak at
the core center and is consistent with experiments. Unlike the
case for a pure DSC, there exists no linear relationship between
zero energy LDOS and $T_1^{-1}$. We show that SDW order could
strongly enhance the NMR signal rate, and as a result, the NMR
signal $T_1^{-1}$ from $^{17}O$ site has its largest value at the
core center V, and smallest value at the saddle point S site (see
Figs. 6 and 7). The temperature dependent of $T_1^{-1}$ at  V site
exhibits a sharp peak at low temperature which is in agreement
with the NMR experiments~\cite{NMR,NMR3}. Since the strength of
the SDW order depends on the magnitude of B, the value of
$T_1^{-1}$ should also be enhanced by increasing B. We also
compare $T_1^{-1}$ from $^{63}Cu$ and $^{17}O$ nuclei at different
T and B, their essential features are similar to each other. The
magnitude of $^{17}T_1^{-1}$ outside the vortex core region is
about one order of magnitude smaller than that of $^{63}T_1^{-1}$
while these two values are closer to each other inside the vortex
core. Finally, we would like to  point out that the difference
between our results and those of a similar work~\cite{Taki02} has
been discussed in detail in Section III. Another fundamental
difference is that their LDOS~\cite{Taki02} at the vortex center
for a pure DSC ($U=0$) shows double peaks near E=0 while only a
single peak is obtained in previous~\cite{Wang95}and present
works. It is our hope that the present calculation may shed more
light on the understanding of recent NMR and STM experiments for
the mixed state of HTS in a unified picture, and stimulate more
theoretical activity in this field.

\begin{acknowledgments}
We are grateful to S. H. Pan, B. Friedman, G. -q. Zheng and Z. D.
Wang for useful discussions. This work was supported by a grant
from the Robert A. Welch Foundation and by the Texas Center for
Superconductivity at the University of Houston through the State
of Texas (YC and CST), and the Department of Energy (JXZ).
\end{acknowledgments}


\begin{thebibliography}{99}
\bibitem{Lake01} B. Lake, G. Aeppli, K.N. Clausen, D.F. McMorrow, K. Lefmann,
N.E. Hussey, N. Mangkorntong, M. Nohara, H. Takagi, T.E. Mason,
and A. Schr\"oder, Science {\bf 291}, 1759 (2001); B. Lake, H.M.
Ronnow, N.B. Christensen, G. Aeppli, K. Lefmann, D.F. McMorrow, P.
Vorderwisch, P. Smeibidl, N. Mangkorntong, T. Sasagawa, M. Nohara,
H. Takagi, and T.E. Mason, Nature {\bf 415}, 299 (2002).
\bibitem{Miller} R.I. Miller, R.F. Kiefl, J.H. Brewer, J.E. Sonier, J. Chakhalian,
S. Dunsiger, G.D. Morris, A.N. Price, D.A. Bonn, W.H. Hardy, and
R. Liang, Phys. Rev. Lett. {\bf 88}, 137002 (2002).
\bibitem{NMR} V.F. Mitrovic, E.E. Sigmund, M. Eschrig,
H.N. Bachman, W.P. Halperin, A.P. Reyes, P. Kuhns, and W.G.
Moulton, Nature {\bf 413}, 501 (2001); V.F. Mitrovic, E.E.
Sigmund, W.P. Halperin, A.P. Reyes, P. Kuhns, and W.G. Moulton, cond-mat/0202368 (unpublished).
\bibitem{NMR1} N.J. Curro, C. Milling, J. Hasse, and C.P. Slichter, Phys. Rev. B {\bf 62}, 3473 (2000).
\bibitem{NMR2} K. Kakuyanagi, K. Kumagai, and Y. Matsuda, Phys. Rev. B {\bf 65}, 060503 (2002).
\bibitem{NMR3} K. Kakuyanagi, K. Kumagai, Y. Matsuda, and M. Hasegawa, cond-mat/0206362 (unpublished).
\bibitem{Arovas97} S.-C. Zhang, Science {\bf 275}, 1089 (1997); D.P. Arovas, A. J. Berlinsky,
 C. Kallin, and Shou-Cheng Zhang, Phys. Rev. Lett. {\bf 79}, 2871 (1997).
\bibitem{Demler} E. Demler, S. Sachdev, and Y. Zhang, Phys. Rev. Lett. {\bf 87}, 067202 (2001);
Y. Zhang, E. Demler, and S. Sachdev, Phys. Rev. B {\bf 66}, 094501 (2002).
\bibitem{Ogata99} M. Ogata, Int. J. Mod. Phys. B {\bf 13}, 3560 (1999).
\bibitem{Machida} M. Ichioka, M. Takigawa and K. Machida, J. Phys. Soc. Jpn. {\bf 70}, 33 (2001).
\bibitem{Zhu01} Jian-Xin Zhu and C.S. Ting, Phys. Rev. Lett. {\bf 87}, 147002 (2001).
\bibitem{chen01} Yan Chen and C.S. Ting, Phys. Rev. B {\bf 65}, R180513 (2002).
\bibitem{Zhu02} Jian-Xin Zhu, Ivar Martin, and A.R. Bishop,
Phys. Rev. Lett. {\bf 89}, 067003 (2002).
\bibitem{Chen02} Yan Chen, H.Y. Chen and C.S. Ting, Phys. Rev. B {\bf 66}, 104501 (2002).
\bibitem{DChen} Han-Dong Chen, Jiang-Ping Hu, Sylvain Capponi, Enrico Arrigoni, and Shou-Cheng Zhang,
Phys. Rev. Lett. {\bf 89}, 137004 (2002).
\bibitem{Franz} M. Franz, D.E. Sheehy, and Z. Tesanovic, Phys. Rev. Lett. {\bf 88}, 257005 (2002).
\bibitem{Taki99} M. Takigawa, M. Ichioka, and K. Machida, Phys. Rev. Lett. {\bf 83}, 3057 (1999);
J. Phys Soc. Jpn. {\bf 69}, 3943 (2000).
\bibitem{wortis} R.Wortis, A.J. Berlinsky, and C. Kallin, Phys. Rev. B {\bf 61}, 12342 (2000).
\bibitem{Morr} D.K. Morr and R. Wortis, Phys. Rev. B {\bf 61}, R882 (2000),
and D.K. Morr, Phys. Rev. B {\bf 63}, 214509 (2001).
\bibitem{Shastry} B.S. Shastry, Phys. Rev. Lett. {\bf 63}, 1288 (1989);
A.J. Millis, H. Monien, and D. Pines, Phys. Rev. B {\bf 42}, 167
(1990); Z.Y. Weng, D.N. Sheng, and C.S. Ting, Phys. Rev. B {\bf
59}, 11367 (1999).
\bibitem{Bulut} N. Bulut, D. Hone, D.J. Scalapino, and N.E. Bickers, Phys. Rev. Lett.
{\bf 64}, 2723-2726 (1990).
\bibitem{chenAPS} Some preliminary results have been reported in,
C.S. Ting, Yan Chen, and Jian-Xin Zhu, Bull. Am. Phys. Soc. {\bf 47} (1), 330 (2002) [March].
\bibitem{Renner} Ch. Renner, B. Revaz, K. Kadowaki, I. Maggio-Aprile,
\O. Fischer, Phys. Rev. Lett. {\bf 80}, 3606 (1998).
\bibitem{Pan} S.H. Pan, E.W. Hudson, A.K. Gupta, K.-W. Ng, H. Eisaki,
S. Uchida, and J.C. Davis, Phys. Rev. Lett. {\bf 85}, 1536 (2000).
\bibitem{Leadon} R. Leadon, and H. Suhl, Phys. Rev. {\bf 165}, 596 (1968).
\bibitem{Wang95} Y. Wang and A.H. MacDonald, Phys. Rev. B {\bf 52}, R3876 (1995).
\bibitem{Taki91} M. Takigawa, A.P. Reyes, P.C. Hammel, J.D. Thompson, R.H. Heffner,
Z. Fisk, and K.C. Ott, Phys. Rev. B {\bf 43}, 247 (1991).
\bibitem{Bulut92} N. Bulut and D.J. Scalapino, Phys. Rev. B {\bf 45}, 2371 (1992).
\bibitem{smallangle} R. Gilardi, J. Mesot, A. Drew, U. Divakar, S.L. Lee,
E.M. Forgan, O. Zaharko, K. Conder, V.K. Aswal, C.D. Dewhurst, R. Cubitt, N. Momono,
and M. Oda, Phys. Rev. Lett. {\bf 88}, 217003 (2002).
\bibitem{Knapp} D. Knapp, C. Kallin, A.J. Berlinsky, and R.
Wortis, Phys. Rev. B {\bf 66}, 144508 (2002).
\bibitem{Taki02} M. Takigawa, M. Ichioka, and K. Machida, Phys. Rev. Lett. {\bf 90}, 047001 (2003).
\end{thebibliography}
\end{document}